# Geographic Patterns in I2P Peer Selection: An Empirical Network Topology Analysis

**Siddique Abubakr Muntaka** | **Jess Kropczynski** | **Jacques Bou Abdo** | **Murat Ozer**

[1]School of Information Technology, University of Cincinnati, OH, USA

**Correspondence**
Corresponding author Siddique Abubakr Muntaka. Email: muntaksr@mail.uc.edu

**Present address**
2610 University Cir, Cincinnati OH 45221.

## Abstract

The Invisible Internet Project (I2P) routes data via encrypted, decentralized tunnels. Peer selection can significantly affect security and performance. This empirical study examines whether geographic location systematically influences I2P's routing topology. Consistent with I2P's design principles, which include avoiding multiple peers from the same /16 IP subnet to maximize anonymity, we conducted assortativity analysis, community detection, and permutation testing on data from 327 routers and 254 connections (SWARM-I2P). We found a network-level absence of significant geographic homophily. The assortativity coefficient was $r = 0.017$ ($p = 0.222$). Same-country connections (11.1%) are statistically near random expectation (10.91%). Community detection found 110 highly modular groups ($Q = 0.972$) only moderately aligned geographically (NMI = 0.521). We conclude that aggregate peer selection in I2P leads to a highly heterogeneous, random geographical mixing, providing a foundation for understanding the performance-anonymity trade-off.

## 1 | INTRODUCTION

Anonymous networks balance a critical trade-off: geographic proximity improves performance but can create patterns that undermine anonymity[1]. The Invisible Internet Project (I2P) uses a fully decentralized, peer-to-peer architecture with encrypted tunnels instead of centralized directories as in The Onion Router (Tor)[2]. Routers build multi-hop paths using garlic routing by bundling messages with layered encryption to serve users who need strong privacy, such as journalists and activists[3,4]. The terms "node," "peer," and "router" are used interchangeably in this study to refer to participating routers in the I2P network.

A new router joins I2P by contacting reseed servers for an initial peer list (Figure 1), then queries the distributed NetDB managed by floodfill routers. Using Kademlia's XOR metric, it builds exploratory tunnels to discover more peers[5].

Each router maintains a local performance profile. When building a data tunnel, it selects reliable, high-capacity peers while avoiding multiple peers from the same /16 IP subnet for security[6,7].

How peers are selected is fundamental. I2P developers explicitly ask whether using GeoIP to prioritize nearby peers would harm anonymity[8]. Current implementations weight peers by performance metrics like bandwidth[9], but whether geographic proximity systematically influences selection remains unknown. If performance weighting creates geographic clustering, regional adversaries could exploit it[10]. Conversely, enforcing full geographic randomness may increase latency. We need an empirical baseline; therefore, we need to determine whether I2P is already clustered or randomly mixed.

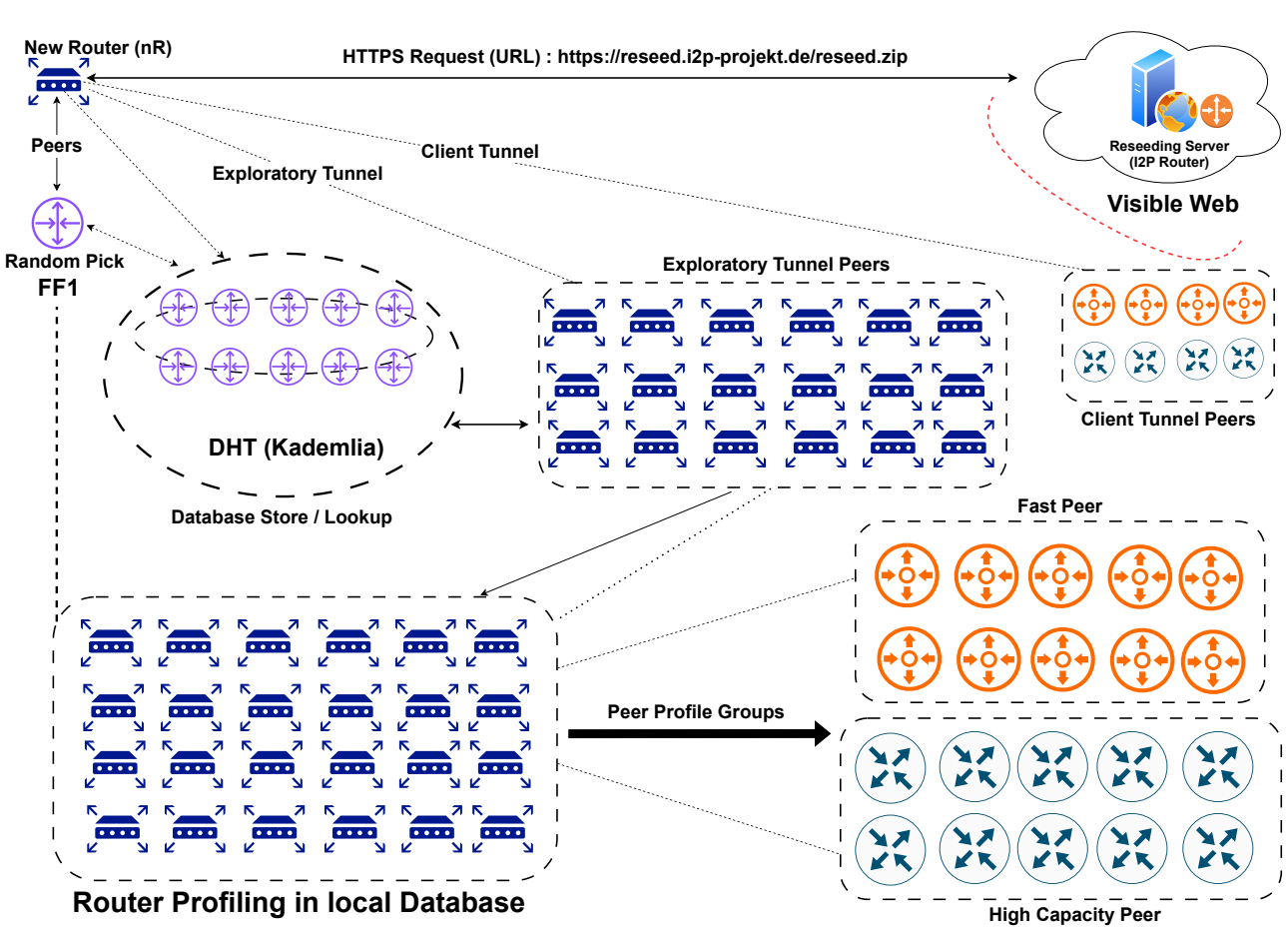

**FIGURE 1** The process of a new I2P router joining the network and forming tunnels.

We address this gap using social network analysis to answer: (1) whether I2P exhibits network-level geographic homophily, (2) how topological communities align with geography, and (3) if local router profiling explains global patterns.

Our contributions include the first empirical measurement of geographic homophily in I2P, analysis of community-geography alignment, and examination of local-to-global dynamics.

The remainder of this study is organized as follows: Section 2 presents the literature related to the topic; Section 3 elaborates on the methodology; Section 4 presents the research findings; Section 5 discusses the implications of the research; and Section 6 provides concluding remarks.

## 2 | RELATED WORK

I2P's architecture differs fundamentally from networks like Tor. While Tor uses a directory-based system, I2P is fully decentralized and peer-to-peer [11]. All routers function as both clients and servers, building temporary encrypted tunnels [12]. Its core technique, called garlic routing, helps bundle messages into encrypted packets with layered routing instructions [13]. Tunnels are unidirectional and short-lived, rebuilt every few minutes (usually ten). Thus, two-way communication requires a pair of tunnels: one for outgoing requests and another for incoming replies [14], as shown in Figure 2.

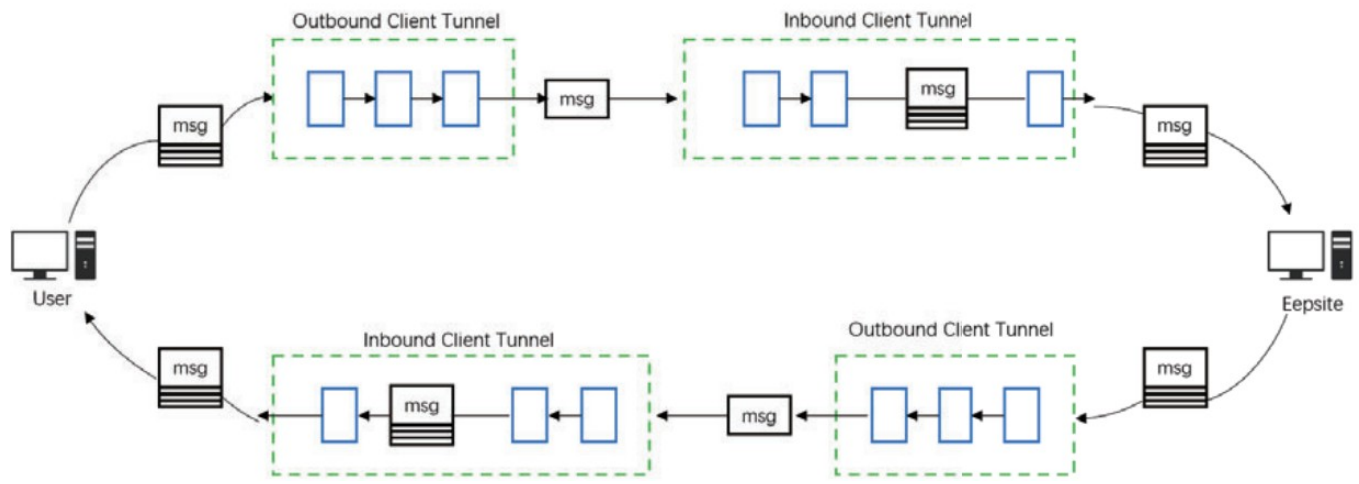

**FIGURE 2** Schematic of I2P Eepsite communication [14].

The network operates via a distributed database (NetDB) based on a Kademlia distributed hash table (DHT) [5]. Specialized floodfill routers manage this database, eliminating single points of failure but making network measurement challenging [15]. Consequently, each router maintains a unique, personalized network view built through peer interactions and NetDB queries to floodfill routers [4].

This decentralized nature has complicated I2P measurement. Early studies estimated about 25,000 daily active peers, though coverage was incomplete [16]. Hoang et al. [10] suggested a stable size near 32,000 peers daily, with concentrations in North America, Russia, and Western Europe. While some research indicates that a router's location can largely influence peer discovery [17], a critical gap persists in peer selection for tunnels. It remains unknown whether this geographic effect extends to the final selection of peers for tunnel construction. This question is also highlighted by I2P developers [8].

Tor research has more thoroughly examined geography's role, studying how relay distribution may affect anonymity [18] and whether attacks like website fingerprinting vary by region [19]. However, Tor's fixed directory and relay model differ sharply from I2P's fluid nature for peer-to-peer design. Within I2P research, focus has been on applications and services like file sharing and eepsites (I2P sites, a name similar to website in visible web) [15] or security analysis, leaving the network-level geographic influence on peer selection understudied.

Our work addresses this gap using established network science methods. We apply assortativity analysis to measure homophily, the tendency for nodes with similar country attributes to connect [20]. We also use community detection algorithms, such as the Louvain method, to identify densely connected groups [21]. While effective for analyzing distributed systems [22], these techniques have seen limited application to empirical I2P tunnel data due to challenges in data collection. Our research applies these methods to a novel I2P dataset [4] and systematically investigates geographic patterns in peer selection.

## 3 | METHODOLOGY

We adopted an empirical approach to detect geographic patterns in I2P peer selection. Data collection involved passive monitoring of the live network, followed by network graph construction and analysis through social network methods.

### 3.1 | Data Collection and Network Construction

Data were gathered from the operational I2P network over three days (April 16-18, 2025) using the SWARM-I2P framework [4]. This three-day observation period captured 327 routers across 39 countries, providing sufficient sample size for network-level geographic analysis. This deployed a standard I2P router configured for passive monitoring, recording connection metadata without intercepting application data.

Observations came from four sources: (1) metadata when our router served as a transit node in others' tunnels; (2) paths of tunnels our own router built; (3) queries to the local NetDB and floodfill routers for peer identification; and (4) local performance profiles tracking bandwidth and uptime.

Processing aggregated 3,883 observed connections into 254 unique directed edges, weighted by frequency. From 788 router records, deduplication by cryptographic hash yielded 327 unique nodes. GeoIP lookups attributed countries to nodes across 39 nations, though 20.8% remained unknown (likely VPN or advanced privacy users). Notable concentrations were in the Russian Federation (28.4%), United States (25.7%), and Germany (16.8%). We defined the full network as directed graph $G = (V, E)$ with $|V| = 327$, $|E| = 254$. For geographic analysis, we defined subgraph $G_{geo}$ consisting of the 226 directed edges where both source and target nodes had known country attributions, connecting 250 nodes across 39 countries.

## 3.2 | Analytical Framework

We employed three complementary techniques. First, we calculate the categorical assortativity coefficient $r$ (Equation 1) to measure same-country connection tendency. Values near zero indicate random mixing; positive values indicate homophily, as supported in Yuan et al[23]. We considered $r > 0.3$ to be substantial homophily, following the landmark work in[20].

$$r = \frac{\sum_i e_{ii} - \sum_i a_i b_i}{1 - \sum_i a_i b_i} \tag{1}$$

where $e_{ii}$ is the fraction of edges connecting nodes in category $i$ (country) to other nodes in the same category, $a_i$ is the fraction of edges originating from nodes in category $i$, and $b_i$ is the fraction of edges terminating at nodes in category $i$.

Second, we performed permutation testing with 1,000 iterations to assess statistical significance, shuffling country labels while preserving topology. We also compared observed same-country connection rates to a random baseline.

Third, we applied the Louvain algorithm for community detection[21] and evaluated alignment with geography using Normalized Mutual Information (NMI)[24]. Finally, we computed a geographic diversity score for each community $C$: unique countries in $C$ divided by total routers in $C$. A mean score near 1.0 indicates geographically mixed communities.

# 4 | RESULTS

Our analysis of the I2P network graph reveals a topology characterized by geographic randomness at the network level, despite strong internal structure. The network, $G$, comprised of 327 routers connected by 254 directed tunnels, forms a sparse graph with a density of 0.0024 and low reciprocity (0.024), consistent with I2P's unidirectional architecture; since inbound and outbound tunnels are constructed independently through distinct peer sets, bidirectional observation of the same peer pair is inherently rare rather than a measurement artifact. The network fragmented into 77 components, with the largest containing 135 nodes. For geographic analysis, we focused on subgraph $G_{geo}$, containing 250 nodes with known country attributions and 226 connections.

## 4.1 | Geographic Mixing Patterns

The connection patterns immediately suggested limited geographic bias. Of the 254 total edges, the analysis focuses on the 226 edges that connect nodes with known locations. Within this subset, cross-country connections dominated, comprising 201 edges (88.9%), while only 25 edges (11.1%) connected peers within the same country (Figure 3 left).

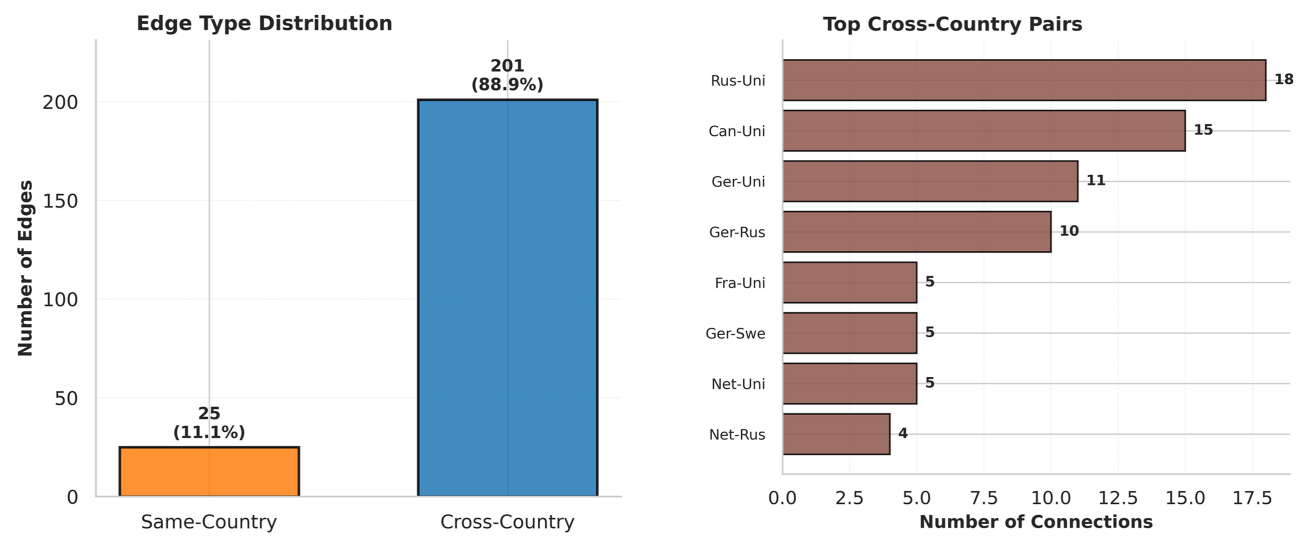

**FIGURE 3** Geographic connection patterns. (Left) Edge distribution shows 11.1% same-country connections. (Right) Top cross-country connection pairs.

The distribution of same-country connections (Table 1) showed activity concentrated in the United States (10 edges), the Russian Federation (7), and Germany (5). More revealing were the frequent long-distance connections (Figure 3 right), with Russian Federation-United States (18 edges) and Germany-United States (11 edges) pairs occurring regularly, suggesting factors beyond proximity drive peer selection.

**TABLE 1** Top Same-Country Connections

| Country | Same-Country Edges | % of Country's Edges |
|---|---|---|
| United States | 10 | 11.2% |
| Russian Federation | 7 | 9.2% |
| Germany | 5 | 12.2% |
| Netherlands | 2 | 7.7% |
| France | 1 | 6.2% |

## 4.2 | Statistical Evidence of Random Mixing

Formal assortativity analysis confirmed these observations. The categorical assortativity coefficient for country was $r = 0.0170$, exceptionally close to zero and well below thresholds for meaningful homophily. Permutation testing with 1,000 iterations generated a null distribution (mean = –0.0031, std dev = 0.0261) that contained our observed value, yielding a statistically non-significant p-value of 0.2220 (Figure 4 left). This fails to reject the null hypothesis of random geographic mixing.

As independent validation, we compared observed connection rates to theoretical random expectation. Given the geographic distribution of nodes, the expected rate of same-country connections was 10.91%. Our empirical data showed an almost identical rate of 11.06%, yielding an observed-to-expected ratio of 1.01 (Figure 4 right). This near-perfect correspondence provides strong evidence that geographic proximity does not systematically bias peer selection in I2P.

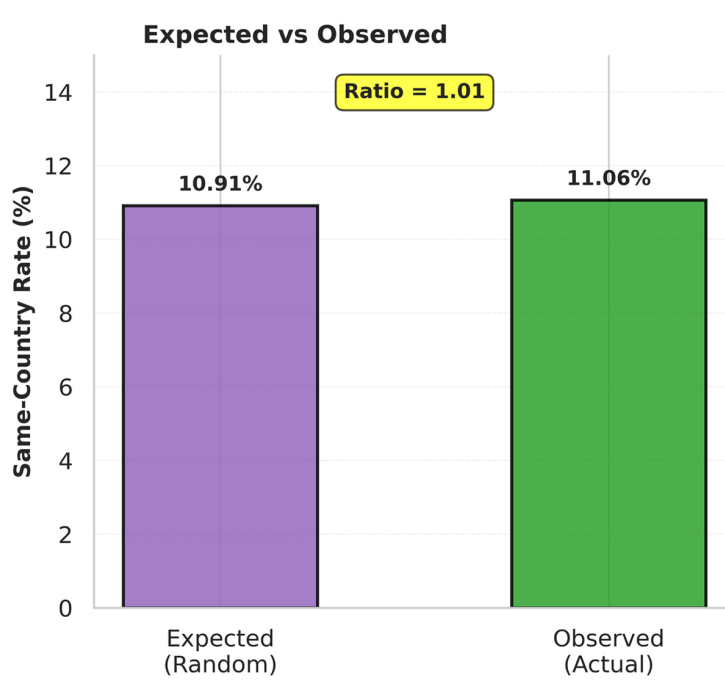

**FIGURE 4** Statistical validation of random geographic mixing. (Left) Permutation test shows $r = 0.0170$ falls within null distribution ($p = 0.2220$). (Right) Observed same-country rate (11.06%) matches random expectation (10.91%).

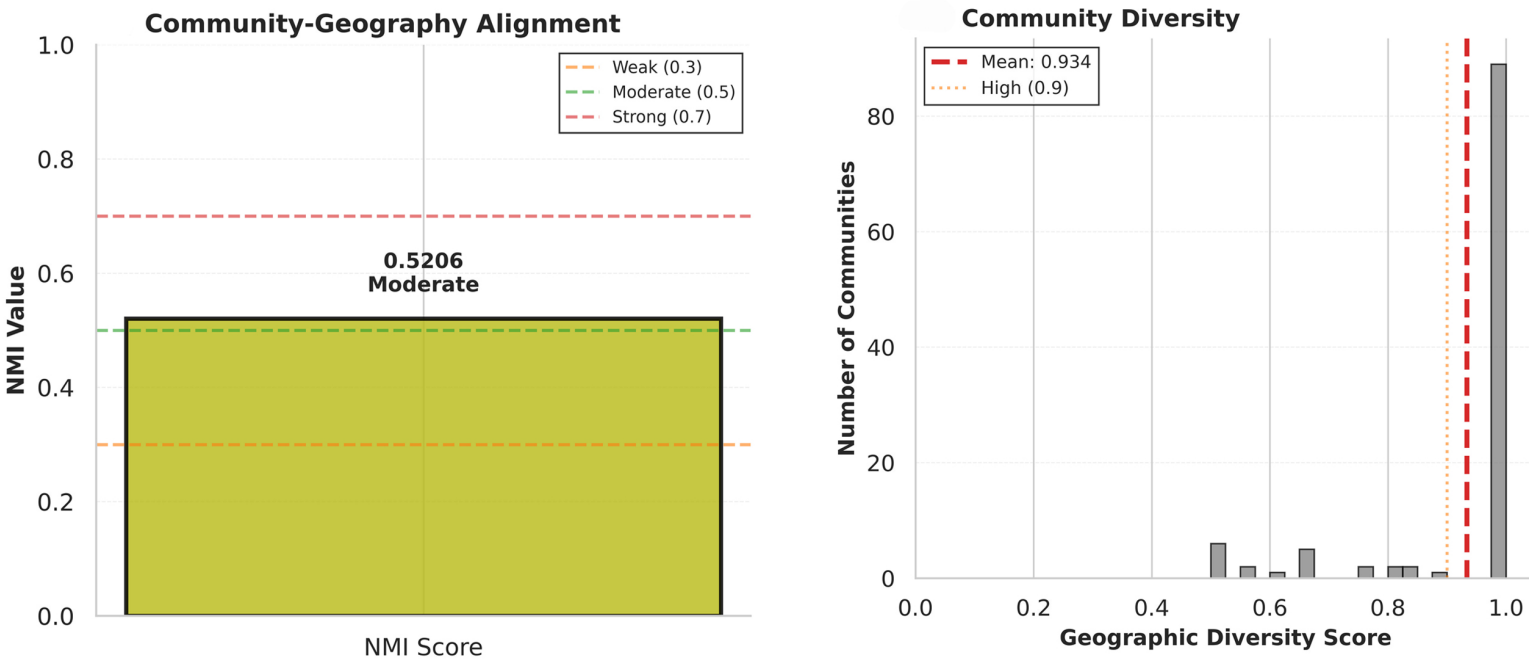

**FIGURE 5** Community structure and geographic diversity. (Left) NMI score of 0.5206 shows moderate community-geography alignment. (Right) Mean diversity score of 0.934 indicates highly mixed communities.

## 4.3 | Community Structure and Geographic Diversity

Community detection revealed strong topological organization with geographic heterogeneity. The Louvain algorithm identified 110 communities with very high modularity ($Q = 0.9717$), a score partly expected given the network's sparse, fragmented topology (77 components), indicating strong internal organisation. However, these communities showed only moderate geographic alignment (NMI = 0.5206), suggesting that country is one of several factors influencing community formation (Figure 5 left). Given the I2P peer selection mechanism that prioritizes local performance metrics, this high modularity strongly indicates communities clustering around non-geographic attributes, such as sustained bandwidth, uptime, or stability metrics recorded in local performance profiles.

The geographic diversity within communities was remarkable. The mean diversity score across all 110 communities was 0.934, indicating most communities contain routers from nearly as many countries as they have members (Figure 5 right). A full 80.9% of communities exhibited perfect diversity (score = 1.0); since no community scored strictly between 0.9 and 1.0, the High Diversity (>0.9) category yields an identical count (Table 2). These categories are non-exclusive: the diversity metric assigns a score of 1.0 to singleton communities, meaning isolated single-router communities appear in both Perfect Diversity and Single Country Origin simultaneously, though 38.2% consisted of single-country nodes, primarily isolated pairs or singletons. We observed a weak negative correlation between community size and diversity ($r$ = –0.456), suggesting larger clusters incorporate slightly less geographic mixing while remaining highly diverse (Figure 6).

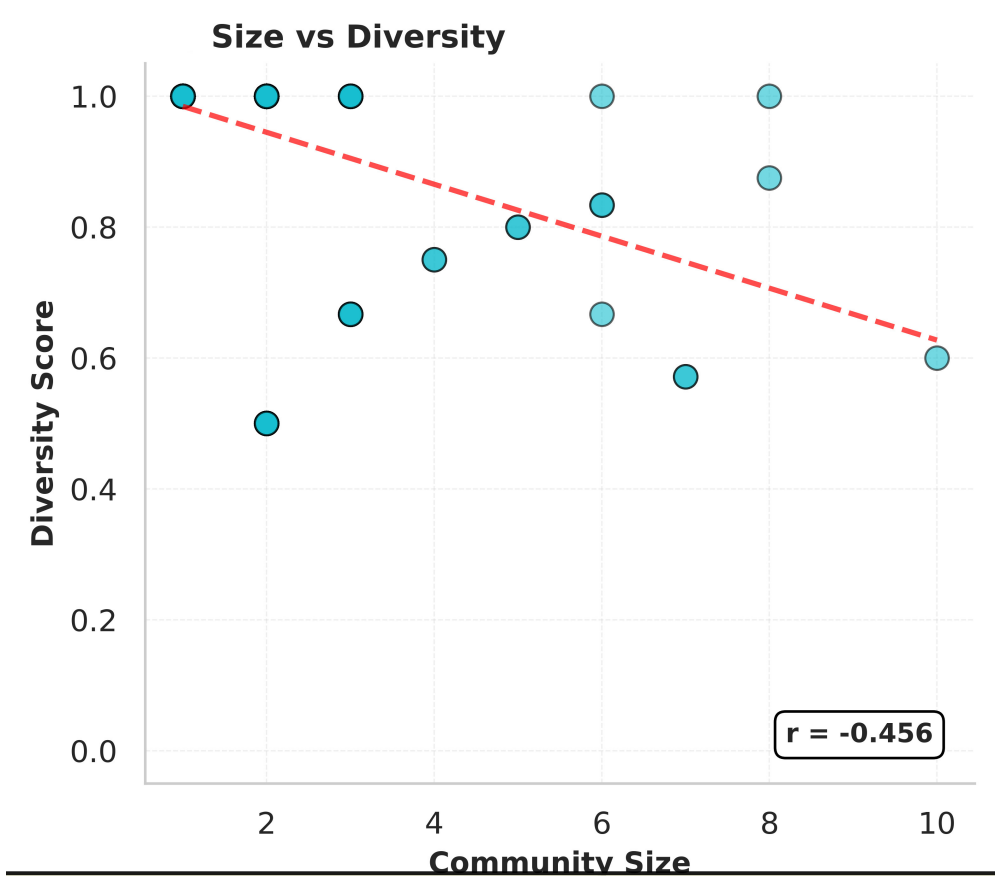

**FIGURE 6** Weak negative correlation ($r$ = –0.456) between community size (number of routers) and diversity.

**TABLE 2** Geographic Diversity Distribution

| Diversity Category | Communities | Percentage |
|---|---|---|
| Perfect Diversity (1.0) | 89 | 80.9% |
| High Diversity (> 0.9) | 89 | 80.9% |
| Single Country Origin | 42 | 38.2% |

## 4.4 | Synthesis of Findings

Four complementary analyses converge on a consistent conclusion (see Table 3): geographic location does not systematically drive peer selection in I2P. The negligible assortativity ($r$ = 0.0170) with statistical non-significance ($p$ = 0.222), the nearly identical observed and expected same-country connection rates (ratio=1.01), the moderate community-geography alignment (NMI=0.5206), and the exceptional internal diversity of communities (mean=0.934) collectively demonstrate that I2P's peer selection achieves essentially random geographic mixing at the network level, despite strong topological structure.

**TABLE 3** Summary of Key Statistical Results

| Metric | Value | Interpretation |
|---|---|---|
| Assortativity $r$ | 0.0170 | Very weak |
| Permutation $p$-value | 0.222 | Not significant |
| Obs/Exp ratio | 1.01 | Consistent with random |
| Modularity $Q$ | 0.9717 | Very strong structure |
| NMI | 0.5206 | Moderate alignment |
| Mean Diversity | 0.934 | Very high mixing |

# 5 | DISCUSSION AND CONCLUSION

This study clarifies a key empirical fact: peer selection in the I2P network is not associated with apparent systematic geographic bias at the aggregate level. Using our quantitative analysis, we show that assortativity is essentially null ($r$ = 0.017), that observed connection rates are consistent with chance, and that the identified communities are highly heterogeneous. This convergence across four independent analytical methods, each measuring a distinct topological dimension from a single-router vantage point inherent to I2P's decentralized architecture, provides strong evidence that the observed geographic randomness is a genuine emergent property of I2P's peer selection rather than an observational artifact.

This baseline measurement addresses a critical design question explicitly raised by I2P developers: whether geographic proximity should be prioritized in peer selection. Previous studies focused on network size estimation or application-level analysis but lacked empirical measurement of peer selection patterns due to the challenges of observing I2P's decentralized topology. By establishing that geographic randomness emerges naturally from I2P's current design, our findings provide the empirical foundation necessary for future optimization efforts. Network designers can now evaluate proposed performance improvements (e.g., geographic proximity preferences) against this baseline to quantify their trade-offs with anonymity properties, while security analysts have a verified understanding of whether exploitable geographic clustering exists.

The observed network structure, marked by very high modularity ($Q$ = 0.9717) that is only moderately aligned with geography ($NMI$ = 0.5206), suggests a dual mechanism at play. Specifically, local optimization mechanisms likely drive the strong internal connectivity (modularity) based on router performance profiles (e.g., bandwidth and uptime), while the system's security constraints (e.g., avoiding multiple peers from the same /16 IP subnet) are the most plausible mechanism to prevent this performance-based clustering from collapsing into geographic homophily. Though local optimisation mechanisms are in effect, and high network churn may contribute to peer diversity, geographic randomness is an emergent collective property;

since I2P peer selection is governed by performance metrics and /16 IP subnet constraints rather than GeoIP data, the routing mechanism cannot produce geographic bias from information it does not possess, regardless of how the 20.8% of unattributed nodes are physically distributed. As a result, I2P's decentralized architecture results in a random geographic distribution. Although individual routers can favourably connect to proximate high-performance peers, aggregating hundreds of autonomous routing choices, selectively filtered by diverse peer pools and sustained by dynamic churn, leads to a globally heterogeneous topology. This inherent randomness underpins anonymity properties, increasing the cost of surveillance in a region and making routing paths less predictable. Consequently, I2P currently trades potential latency gains for increased anonymity. It is the duty of any possible optimisation to protect this randomness through imposing diversity constraints. The current research provides the necessary starting point for the latter kind of assessment. Also, the study supports the effectiveness of the I2P design in achieving its primary privacy goal: realizing its geographically neutral peer-selection protocol.

## 6 | LIMITATIONS AND FUTURE WORK

The study has limitations that outline clear directions for future research. The research is based on a static observation of 327 router nodes; this is statistically sound, but the sample is only a subset of the network. Country-level attribution can obscure patterns at the city/autonomous system (AS) level, and the passive observation technique records only successful relationships from a single vantage point. Since peer availability in decentralized systems is relative to the observer's network position, this study presents a valid localized topology, acknowledging that distinct geographic vantage points may reveal different peer horizons. These limitations can be overcome in future studies by using longitudinal measurement to determine the temporal stability. Multi-scale geographic analysis is also justified to determine whether randomness occurs at more minor scales. Furthermore, active probing and multi-point observation from geographically diverse vantage points are necessary to gather a more detailed view of the selection process and improve external validity. Empirical experimentation alongside source-code inspection can be key to confirming the operation mechanism and quantifying the definite performance cost of the current random-mixing paradigm. Further future research could also involve comparative geographic patterns of I2P deployments, cross-network studies using Tor or Freenet, and attempts to isolate the architectural effect. Lastly, adversarial modelling might be used to emulate even small-scale geographic biases on susceptibility to traffic correlation or Sybil attacks.